# *Molecular electronics at Metal / Semiconductor Junctions*
# Si inversion by Sub-nm Molecular Films

**Omer Yaffe, Luc Scheres [1], Sreenivasa Reddy Puniredd [2], Nir Stein, Ariel Biller, Rotem Har Lavan, Hagay Shpaisman, Han Zuilhof,[1] Hossam Haick,[2] David Cahen\* and Ayelet Vilan\***

*Dept. of Materials & Interfaces, Weizmann Inst. of Science, Rehovot 76100, ISRAEL*

[1] *Lab. of Organic Chemistry, Wageningen Univ., Dreijenplein 8,*

*6703 HB Wageningen, the Netherlands*

[2] *Dept. Chemical Engineering and Russell Berrie Nanotechnology Inst., Technion - Israel Inst. Technology, Haifa 32000, Israel*

**Abstract**     Electronic transport across n-Si-alkyl monolayer/Hg junctions is, at reverse and low forward bias, independent of alkyl chain-length from 18 *down to 1 or 2 carbons!* This and further recent results indicate that electron transport is *minority*, rather than *majority* carrier-dominated, occurs via generation and recombination, rather than (the earlier assumed) thermionic emission and, as such is rather insensitive to interface properties. The (m)ethyl results show that binding organic molecules directly to semiconductors provides semiconductor/metal interface control options, not accessible otherwise.

Using a semiconductor, especially Si, instead of a metal, as one of the electrodes to contact the molecules not only adds another "knob" to turn to study transport through molecules, but it also can make molecular electronic junctions more relevant for potential future electronics. Implicit in most of the increasing efforts in this direction is the assumption that the molecules influence, possibly even control, the junction charge transport characteristics. In this and forthcoming reports we examine and try to define when and in how far this assumption holds. Specifically we use the n-*Si-alkyl chain/Hg* junction, often used in laboratory tests, where Hg contacts a monolayer of $C_nH_{2n+1}$ alkyl chain molecules, bound by direct Si-C bonds to Si <111>[1]. Here we focus on how molecule-semiconductor interaction can dictate the mechanisms of charge transport through these junctions.

Understanding the transport mechanism(s) across molecular junctions is critical for rational design of possible future molecule-based devices. Previously[1] charge transport across junctions with n-Si was described as thermionic emission (TE) at low forward and reverse bias, and tunneling at high forward bias. This model was extended further, to include also p-Si[2]. We present and discuss here new experimental results that point to diffusion and recombination of *minority* carriers, rather than *majority* carrier TE, as the dominant transport mechanisms at low forward and reverse bias in n-*Si-alkyl chain/Hg* junctions. Minority carrier transport also explains earlier experimental results that are not consistent with TE.

In Fig. 1 we show current density-voltage (*J-V*) characteristics of n-*Si-alkyl chain/Hg* junctions, made with (1-10) Ω·cm Si <111>. The length of the alkyl chain $C_nH_{2n+1}$ varies from n = 18 down to 1, which means that the monolayer thickness decreases from 2.2 down to 0.2 nm. Monolayer preparation followed, and characterization agreed with literature descriptions for both 'short' monolayers[3,4] and 'long' ones[5]. J-V curves were measured (see inset to Fig. 1) at room temperature, using a controlled growth hanging Hg (99.9999 % purity) drop (HMD) electrode apparatus (Polish Academy of Sciences) and In-Ga eutectic as back contact, in a controlled environment glove box with 10 % relative humidity.

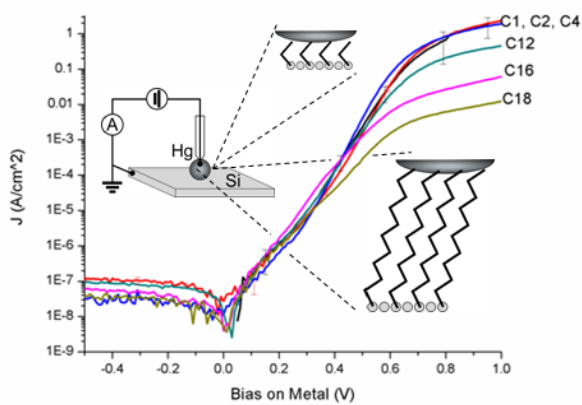

**Figure 1**: J-V curves of n-Si-$C_nH_{2n+1}$/Hg (n=1, 2, 4, 12, 16 and 18). Current densities where calculated using the optically determined nominal contact area of 0.3 mm². Bias is applied to the Hg; Si is grounded. Results are logarithmic averages of at least 7 different junctions with a scan rate of 20 mV/s. Error bars represent standard deviations. The insets give "artist' views of the junctions for 'short' and 'long' monolayers (not to scale!).

Remarkably, the results are independent of molecular length at reverse and low (< 0.4 V) forward bias, *including those obtained with the very short n = 1 and n = 2 molecules.* This result is even more surprising if we consider that the n = 2 coverage is ~50 %, as is the case for all the n ≥ 2 alkyl chains used here, while for n=1 the coverage is roughly double that, ~ 100 %, an issue that is discussed further, below. Molecular (organic insulator) length-dependence enters only at higher forward bias[1], a bias range that will be re-analyzed in a subsequent report. Nevertheless, the exponential current decay with length (at high bias) proves that the molecular monolayer is not shorted electrically.

The reverse and low forward bias length-independence, for C12-C18 monolayers (that correspond to a thickness range of 1.6 - 2.2 nm), was earlier interpreted in terms of a large *Schottky barrier* that creates a charge transport "bottleneck", leading to a negligible effect of differences in monolayer thickness[1]. By adding n = 1, 2 results (Fig.1) the thickness range is significantly increased and, *still*, there is a lack of dependence on molecular length. This result is remarkable because, using a basic tunneling model[6], the tunneling probability through the n=1 monolayer is nine orders of magnitude greater than that through the n=18 monolayer.

In view of these experimental findings we propose that the electrical characteristics of these junctions are best described by diffusion and recombination of *minority* carriers, as was done for MIS tunnel diodes with $SiO_2$ as an insulator[5,7-9]. If the built-in potential ($\psi_{bi}$) at a given semiconductor interface exceeds the energy difference between the actual Fermi level, $E_F$, and the intrinsic Fermi level, $E_i$, in the bulk ($\psi_B$), the region of the semiconductor adjacent to the metal becomes inverted, i.e., there are more minority than majority carriers in that region. If $\psi_{bi} > 2\psi_B$ the surface is in deep inversion[10].

Figure 2 shows a schematic energy band diagram (not to scale) for an inverted n-type MIS tunnel diode. Three basic charge transport processes under forward bias are indicated: (1) majority carrier (electron) TE, attenuated by the tunneling barrier that is introduced by the molecular monolayer, (2) recombination of electrons and holes in the depletion layer (space charge region, SCR) and (3) diffusion of minority carrier (holes) in the bulk, minority carrier injection. Processes (2) and (3) depend mainly on intrinsic[11] semiconductor properties and under specific conditions, are not sensitive to the tunneling barrier, introduced by the insulator[9].

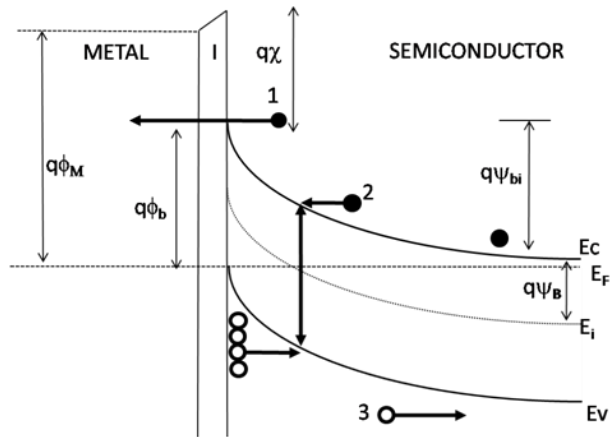

**Figure 2:** Energy-band diagram (not to scale) of inverted metal / n-semiconductor contact with a thin insulating interfacial layer (I). The three basic charge transport mechanisms that are shown are (1) majority carrier thermionic emission, attenuated by tunneling through the organic monolayer, (2) recombination in the depletion region and (3) diffusion of minority carriers in the bulk (with recombination either in the neutral region or at the back contact). The thermionic emission barrier height ($\varphi_b$) is the difference between the metal work function ($\varphi_M$) and the semiconductor electron affinity ($\chi$); ($\varphi_{bi}$) is the built-in potential; ($\psi_B$) is the energy difference between the Fermi level and the intrinsic Fermi level in the bulk; $E_C$, $E_V$ and $E_F$ are the conduction band minimum, the valence band maximum and the Fermi level, respectively; $E_i$ is the intrinsic Fermi level (at the middle of the forbidden energy gap). After ref. [10]

Numerical calculations for charge transport through MIS tunnel diodes (Si-SiO$_2$; SiO$_2$ thickness 1.5-3 nm)[8], showed that for a sufficiently large built-in potential i.e., if the surface is in *deep inversion*, charge transport will be governed *solely by hole injection and generation-recombination* at reverse and low forward bias voltages. In case of very thin insulators, the junction is similar to an abrupt, one-sided p$^+$-n junction[9]. Earlier reverse bias C-V measurements[12,13] that were confirmed by us (see Fig. S1 in supplementary information) indicate that the flat band potential ($\psi_{bi}$) of n-*Si-alkyl chain/Hg* is ~ 0.65 eV. Since the calculated deep inversion potential ($2\psi_B$) of the moderately doped n-Si used here ($N_d$ ~ $10^{15}$ cm$^{-3}$), is ~ 0.6 eV[10], we conclude that the Si surface is indeed in deep inversion at equilibrium[14].

Additional support for minority instead of majority carrier transport at reverse and low forward bias comes from comparing the J-V characteristics for n-Si-$C_{12}$ with Hg or Au as top metal contact (Fig. 3). If an n-semiconductor is in deep inversion, further *increasing* the work-function of the top metal contact will not affect the current at reverse and low forward bias[15]. To test if the n-Si is indeed in deep inversion, we used Au contacts, as Au has 0.2 - 0.5 eV higher work function than Hg[16]. "Ready made" gold Au contacts were used to measure the J-V characteristics[17,18]. The Au contact J-V curve overlaps that obtained with Hg despite the work function difference. The reason is that with the semiconductor in deep inversion, the minority-carrier current is limited by processes in the Si (recombination or diffusion), independent of the band-bending or the presence of an interfacial insulator[9].

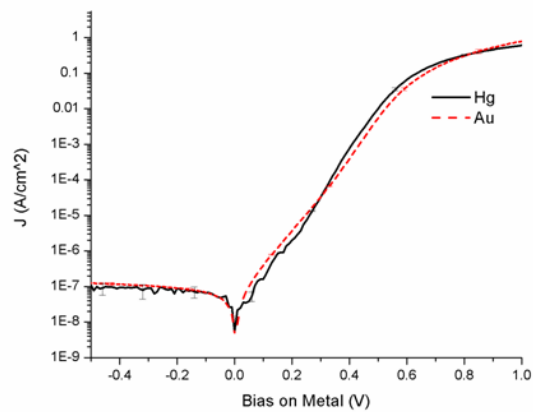

**Figure 3**: Comparison of junctions with Hg and Au contacts to n-Si-$C_{12}H_{25}$. Au contacts were "ready-made", using a "soft" deposition, modified lift-off, float-on method[17,18]. See Fig. 2 for more experimental details.

The role of the insulator in the MIS tunnel diode can be a multiple one[19]: (i) suppress majority carrier TE by introducing an additional tunneling barrier; (ii) introduce an asymmetric tunneling barrier and, by that, favor minority carrier controlled charge transport; (iii) decrease the voltage drop on the semiconductor depletion layer; (iv) contribute an additional dipole and by that alter the semiconductor band bending; (v) reduce the surface recombination rates by passivating the surface states, especially those induced by direct metal – semiconductor interactions, i.e., act as buffer between the metal and the semiconductor[20]. The results with methyl and ethyl (n = 1, 2) films, which are ineffective as insulators under our experimental conditions, show that in the case of the Si-C interface, effects (i), (ii) and (iii) are negligible.

As for effect (v), because Hg does not interact electronically with Si[21], we cannot deduce from the Hg result only, if the organic insulating layer passivates metal-Si interactions in general. However, additional, preliminary (and as yet unpublished) results with "ready-made" Au contacts suggest that these molecules do not completely passivate such interactions.

Therefore, we conclude that (iv), the change in electrostatic potential, due to the dipole layer introduced by chemically binding a molecular monolayer (~ 0.38 eV for Si-CH$_3$[22]), suffices to increase the built-in potential ($\psi_{bi}$) to the level where the n-Si is in deep inversion and by that to *make any majority carrier charge transport negligible*[23]. The fact that half coverage of ethyl can have the same effect as full methyl coverage, indicates that the former suffices for deep inversion and that the additional dipole that likely results with the latter does not change this further.

In principle, close to ideal MIS-like behavior would be possible if we could passivate the Si surface with a "Si oxide monolayer". However, in practice this has not been possible till now, while our results clearly show that even half ethyl-coverage / half H-termination or full methyl-coverage can provide sufficient electrostatic change to passivate the Si and, in this case, to invert it, without the need for a (thicker) insulating layer.

Minority carrier-controlled transport clarifies two previous experimental observations, which seemed inconsistent with (majority carrier) TE. The first one is that the ~ 0.85 eV (effective) TE barrier height, derived from either room temperature J-V or reverse bias C-V ($\phi_b = \psi_{bi} + (E_C - E_F)$) measurements[1,2,12], is almost twice the 0.45 eV activation energy, extracted from the temperature dependence of current-voltage characteristics (J-V-T)[24], using, in our experiments, the same set of Si wafers. J-V-T analysis allows experimental extraction of the charge transport activation energy regardless of any *a priori* assumed transport mechanism (via the Arrhenius relation). While the saturation current for TE is proportional to $\exp\left(-\frac{q\phi_b}{kT}\right)$, where q is the elementary charge, k is the Boltzmann constant, T is the absolute temperature and $\phi_b$ is the barrier height, the saturation current for recombination is proportional to $\exp\left(-\frac{E_g}{2kT}\right)$, where $E_g$ is the semiconductor band gap[25]. If we extract a "barrier height" from J-V-T data there is an implicit assumption that the charge transport is TE-controlled. For n-*Si-alkyl chain/Hg junctions,* the experimentally extracted activation energy (~ 0.45 eV[24]) is much closer to half the band gap of Si (0.51 eV) than to the $\phi_b$ value, extracted from either room temperature J-V or C-V measurements, which points to recombination in the depletion region as the dominant charge transport mechanism.

The second result, which can be explained by minority carrier-controlled charge transport, but not by TE, is the variation of J-V-T characteristic of an n-*Si-alkyl chain/Hg* junction upon low energy electron irradiation[24]. Irradiation was found to considerably increase the number of electronic states in the monolayer ('dope' the monolayer), and this is expressed as a considerable net increase in current. Yet, J-V-T analysis showed that the transport activation energy also increases (instead of decreasing to explain the higher current)[24]. With the new understanding that junction transport before irradiation is minority carrier-controlled, we suggest that irradiation changes the equilibrium electronic state of the junction from deep inversion to depletion. In such a case the charge transport mechanism changes from minority to majority carrier-controlled, i.e., to TE, attenuated by tunneling. Although the activation energy for TE is larger than that for recombination, the net current still increases because of the much higher concentration of majority than minority carriers. Irradiation also decreased the ideality factor toward that for ideal TE (= 1), from the originally higher value, which would fit recombination[26]

We conclude that the well-studied n-*Si-alkyl chain/Hg* is in deep inversion at equilibrium (0 V) and at room temperature. Therefore, charge transport at reverse and low forward bias is controlled by minority carrier generation (at reverse bias), and by recombination and diffusion (at low forward bias) rather than by majority carrier thermionic emission. The indications for the minority carrier control are:

- the current is independent of molecular length, at reverse and low forward bias;
- the experimentally measured built-in potential ($\psi_{bi}$) is larger than the minimal potential for deep inversion ($2\psi_B$);
- the current is independent of the work function of the metal electrode (which will hold for any metal with a work function, higher than that of Hg) and
- the activation energy for charge transport is about half the Si band gap.

While interface energetics (built-in potential) can be extracted from C-V measurements, the shape of the J-V curves is independent of interface properties over most of the bias range considered here.

Recognizing inversion is a crucial step towards understanding the basic science of hybrid metal-molecules-semiconductor (MMS) devices and mapping which molecular aspects control device performance under different conditions. We note that realizing that transport is by minority carriers should be especially useful for solar-cell applications[27,28].

**Acknowledgements:**
We thank L. Kronik, L. Segev (WIS), A. Kahn (Princeton), J. Pelz (Ohio state) and R. Tung (CUNY) for fruitful discussions, the Israel Science Foundation, ISF (Jerusalem), through its Converging Technology and Centre of Excellence programs and the Minerva Foundation (Munich) for partial support, and 21 Ventures for a generous research grant. HS holds an ISF convergent technology predoctoral fellowship. HH holds the Horev Chair for leader in Science and Technology and thanks the US-Israel BSF (Jerusalem) for financial support. HS holds an ISF converging technology pre-doctoral fellowship. HZ thanks NanoNed, funded by the Dutch Ministry of Economic Affairs (project WSC.6972) for financial support. DC holds Rowland and Sylvia Schaefer chair of Energy Research.

**Supporting Information Available:** Capacitance-Voltage measurements of varying monolayer thickness. This material is available free of charge via the Internet at http://pubs.acs.org

**Supporting Information for:**

# *Molecular electronics at Metal / Semiconductor Junctions*
## Si inversion by Sub-nm Molecular Films

Omer Yaffe, Luc Scheres [1], Sreenivasa Reddy Puniredd [2], Nir Stein, Ariel Biller, Rotem Har Lavan, Hagay Shpaisman, Han Zuilhof,[1] Hossam Haick,[2] David Cahen and* Ayelet Vilan*

Capacitance-Voltage measurements on n-Si-$C_nH_{2n+1}$/Hg junctions with monolayer thickness varying from n = 1 (0.2 nm) to n = 18 (2.2 nm).

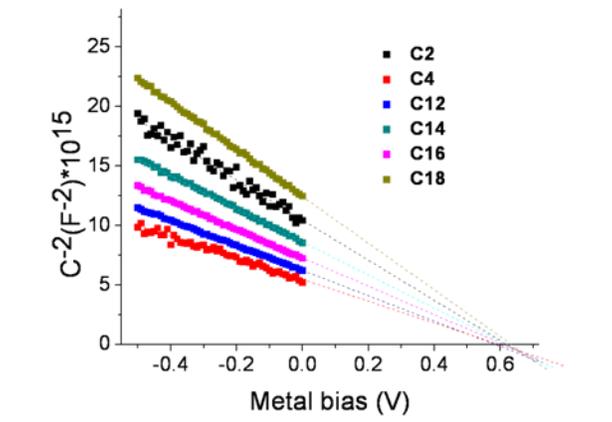

**Figure S1:**

($1/C^2$) - V plots of n-Si-$C_nH_{2n+1}$/Hg (n = 1, 2, 4, 12, 14, 16 and 18) junctions. Dashed lines are extrapolation to ($1/C^2$) = 0, where $\psi_{bi}$ is extracted. The slope is not identical because of small size variation in the Hg drop. Sample preparation is identical to that of Fig. 1. Capacitance was measured with an HP4284A precision LCR meter using a 10 mV AC amplitude at 500 kHz. This frequency was chosen so as not to be too low, to prevent quasi-static behavior and to prevent possible interface states from following the AC signal. The impedance model used for the *C-V* measurements was a parallel circuit of a resistor and capacitor [13] and the contact area for all measurements was $3·10^{-3}$ $cm^2$.

# References

(1) Salomon, A.; Bocking, T.; Seitz, O.; Markus, T.; F., A.; Chan, C.; Zhao, W.; Cahen, D.; Kahn, A.; *Phys. Rev. Lett.* **2005**, *95*, 266807.
(2) Salomon, A.; Boecking, T.; Seitz, O.; Markus, T.; Amy, F.; Chan, C.; Zhao, W.; Cahen, D.; Kahn, A. *Adv. Mater.* **2007**, *19*, 445-450.
(3) Puniredd, S. R.; Assad, O.; Haick, H. *J. Am. Chem. Soc.* **2008**, *130*, 9184-9185.
(4) Puniredd, S. R.; Assad, O.; Haick, H. *J. Am. Chem. Soc.* **2008**, *130*, 13727-13734.
(5) Scheres, L.; Arafat, A.; Zuilhof, H. *Langmuir* **2007**, *23*, 8343-8346.
(6) Simmons, J. G. *J. Appl. Phys.* **1963**, *34*, 1793-1803.
(7) Tarr, N. G.; Pulfrey, D. L.; Camporese, D. S. *IEEE Trans. Electron Devices* **1983**, *30*, 1760.
(8) Shewchun, J.; Green, M. A.; King, F. D. *Solid-State Electron.* **1974**, *17*, 563.
(9) Green, M. A.; King, F. D.; Shewchun, J. *Solid-State Electron.* **1974**, *17*, 551.
(10) Sze, S. M.; NG, K. K. *Physics of Semiconductor Devices*; third ed.; New York: John Wiley & Sons, Inc., 2007.
(11) Surface states can increase the recombination current, an effect that is ignored here, for simplicity's sake.
(12) Maldonado, S.; Plass, K. E.; Knapp, D.; Lewis, N. S. *J. Phys. Chem. C* **2007**, *111*, 17690-17699.
(13) Faber, E. J.; de Smet, L.; Olthuis, W.; Zuilof, H.; Sudholter, E. J.; Bergveld, P.; Van den Berg, A. *ChemPhysChem* **2005**, *6*, 2153-2166.
(14) For devices with a surface potential larger than the inversion potential, the intercept of the $1/C^2$ vs. V curve with the voltage axis (V), is not the flat band potential but the lower limit for the flat band potential. This is because the increase in the depletion layer width with increasing surface potential is minor for inverted surfaces compared to depleted surfaces (Shewchun, loc. cit.)
(15) Camporese, D. S.; Pulfrey, D. L. *J. App.Phys.* **1985**, *57*, 373.
(16) The exact difference depends on how clean the gold is.
(17) Shimizu, K. T.; Fabbri, J. D.; Jelincic, J.; Melosh, N. A. *Adv. Mater.* **2006**, *18*, 1499-1504.
(18) Vilan, A.; Cahen, D. *Adv. Funct. Mater.* **2002**, *12*, 795-807.
(19) Fonash, S. J. *Solar Cell Device Physics*; Academic press: Belton, teaxs, 1981.
(20) Tung, R. T. *Mater. Sci. Eng., R* **2001**, *35*, 1.
(21) Wittmer, M.; Freeouf, J. L. *Phys. Rev. Lett.* **1992**, *69*, 2701.
(22) Ralf, H.; Rainer, F.; Bengt, J.; Wolfram, J.; Lauren, J. W.; Nathan, S. L. *Phys. Rev. B* **2005**, *72*, 045317.
(23) (iv), the electrostatic effect of the adsorbate is naturally also the one that is critical for sensors, such as a CHEMFET (Northrop, loc.cit.) or MOCSER(Cahen et al., loc. cit.)`; those effects are likely to be different for adsorption on bulk and nano sized (e.g., nanowires) substrates, in part because then bulk electrostatic analyses such as those used here, are not or only partially applicable
(24) Seitz, O.; Vilan, A.; Cohen, H.; Haeming, M.; Schoell, A.; Umbach, E.; Kahn, A.; Cahen, D. *Adv. Funct. Mater.* **2008**, *18*, 2102-2113.
(25) Sah, C. T.; Noyce, R. N.; Shockley, W. *Proceedings of the IRE* **1957**, *45*, 1228.
(26) We note that an ideality factor > 1 can also have other reasons, e.g., a tunnel barrier in series with TE(Rhoderick loc.cit)
(27) Har-Lavan, R.; Ron, I.; Thieblemont, F.; Cahen, D. *App. Phys. Lett.* **2009**, *In press*.
(28) Maldonado, S.; Knapp, D.; Lewis, N. S. *J. Am. Chem. Soc.* **2008**, *130*, 3300-3301.
(29) Northop, R. B. *Introduction to Instrumentation and Measurements*; 2nd ed.; CRC press: Boca Raton, 2005.

(30) Cahen, D.; Naaman, R.; Vager, Z. *Adv. Funct. Mater.* **2005**, *15*, 1571-1578.
(31) Rhoderick, E. H. *Monographs in Electrical and Electronic Engineering. Metal-Semiconductor Contacts*; second ed.; Clarendon press: Oxford, 1988.